\documentclass[preprint,showpacs,preprintnumbers,amsmath,amssymb]{revtex4}

\usepackage{graphicx}
\usepackage{dcolumn}
\usepackage{bm}

\begin{document}

\title{Textural transformations in islands on free standing
Smectic C* liquid crystal films}

\author{Jong-Bong Lee\footnote[1]{Present Address: Harvard Medical School,
Department of Biological Chemistry and Molecular Pharmacology,
Boston, MA 02115},
Dmitri Konovalov\footnote[2]{Present Address: Photon Dynamics Inc.,
5970 Optical Court, San Jose, CA 95138} and Robert B. Meyer}

\affiliation{The Martin fisher school of physics, Brandeis
University, Waltham, MA 02454 USA}

\date{\today}
\begin{abstract}
We report on and analyze the textural transformations in islands,
thicker circular domains, floating in very thin free standing
chiral Smectic C* liquid crystal films. As an island is growing,
an initial pure bend texture of the c-director changes into a
reversing spiral at a critical size. Another distinct spiral
texture is induced by changing the boundary condition at the
central point defect in the island. To understand these
transformations from a pure bend island, a linear stability
analysis of the c-director free energy is developed, which
predicts a state diagram for the island. Our observations are
consistent with the theoretical phase diagram.
\end{abstract}

 \maketitle
\section{introduction}
Defects and singularities of the director field in two dimensional
(2D) ordered molecular systems with in-plane orientational order
cause fascinating textures, easily visible by means of
polarized-light microscopy.  In this paper we report a detailed
experimental and theoretical study of a set of simple textural
transformations in a system formed by blowing smoke over a thin
free standing film of chiral smectic C (SmC*) liquid crystal.
Sub-micron smoke particles nucleate islands, circular regions of
added smectic layers, which grow to a certain equilibrium size,
with the smoke particle as a point disclination at the center of
the island. The island maintains azimuthal symmetry during growth,
initially with the c-director field, a unit vector $\hat{c}$,
defined by the projection of the tilted long molecular axis onto
the layer, tangentially oriented.  We find that this initially
stable pure bend texture can transform to two other equilibrium
spiral splay-bend textures, either as the islands grow, or when
they are momentarily perturbed by external forces.  To understand
these transformations among stable or metastable textures, we
develop a linear stability analysis for this general class of
textures, which predicts a state diagram, as a function of island
radius, that agrees well with our observations.

Important precursors to this work include an initial report of
these phenomena\cite{equilibriumsize}, the report of a reversing
spiral texture first discovered in domains at the air/water
interface when the SmC* film is both polar and chiral\cite{polar},
and study of the boojum, a point disclination occurring at the
edge of a circular domain\cite{muzny, maclennan}. The theoretical
analysis of the stability of textures around topological defects
has been investigated for Langmuir monolayers with a tilted phase
and for free standing Sm C films\cite{pettey,rivier,fang,sethna}.
Especially important, K. K. Loh {\it et al.}\cite{loh} pointed out
that there can be more than one stable or metastable texture for a
point defect centered in a circular domain, and described the
reversing spiral texture of a circular domain in a chiral system,
which can be energetically stable or metastable. The remarkable
growth of islands to an equilibrium size, once they are nucleated,
is a consequence of their chirality and their resulting polar
(ferroelectric) symmetry.\cite{equilibriumsize}

\section{Textural transformations of islands}
\begin{figure}
\centering
\includegraphics[width=2.0in]{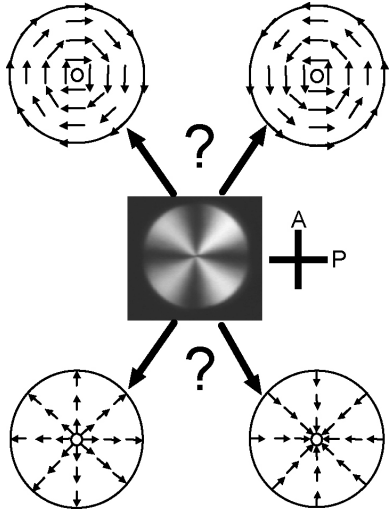}
\caption{Tangential or radial? An island observed with crossed
polarizers and four possible $\hat{c}$ director configurations,
tangential (clockwise or counterclockwise) or radial (outward or
inward).}\label{fig:bend}
\end{figure}

Samples are prepared by drawing a small amount of smectic C*
liquid crystal material across a 6 mm diameter hole in a thin
metal sheet. Materials were prepared from the CS series of
mixtures available from Chisso Co. Typical films as drawn were on
the order of 10 smectic layers thick.\cite{jb}  To create islands,
smoke from burning paper was wafted across the film. Watching the
sample using a polarizing microscope with crossed polarizers, one
could observe the growth of islands nucleated by some of the smoke
particles.  The islands were brighter than the very dark
background film, their brightness being proportional to their
thickness. Once nucleated, islands grew in area but not in
thickness.  As the islands grew, one could observe their internal
texture due to their birefringence. Although sometimes the smoke
particle was the center of a point defect at the periphery of the
island, a so-called boojum, most often the sub-visible smoke
particle was centered in the island as a +1 point disclination, as
shown in Fig.~\ref{fig:bend}.

There are four possible textures for the intensity pattern  shown
in Fig.~\ref{fig:bend}. For a SmC* film of thickness $d$ with
angle $\psi$ between $\hat{c}$ and either of the polarizers, with
crossed polarizers, the intensity of transmitted light is
\begin{equation}\label{transmittedI}
I=I_{0}\sin^{2}2\psi\sin^{2}\Big(\frac{\pi\Delta n d}{\lambda}\Big),
\end{equation}
where $\lambda$ is a wavelength of the incident light and $\Delta
n$ is a birefringence for light propagating normal to the film.

We neglect the effects of the twist of $\hat{c}$ through the film
thickness, since the islands are thin compared to the helical
pitch of the spontaneous twist produced by molecular chirality.
The helical pitch in the materials studied was in the range of 3
to 15 micrometers, while the islands were in the range of
thickness of 0.06 to 0.15 micrometers.\cite{jb} If the $\hat{c}$
director is parallel to either of the polarizers, $\psi=0,\pi$ or
$\pm \pi/2$, then $I=0$. At $\psi=\pm \pi/4$, $\pm 3\pi/4$, $I$ is
a maximum in Eq.~(\ref{transmittedI}). Hence, the four
orientations of $\hat{c}$ indicated in Fig.~\ref{fig:bend} are
possible.

\begin{figure}
\centering
\includegraphics[width=3in]{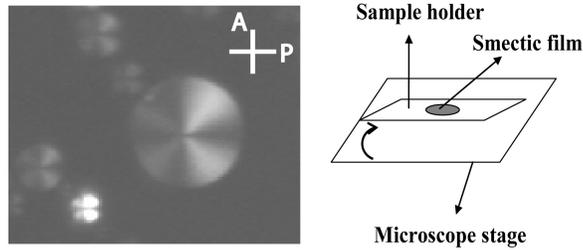}
\caption{Islands observed when the lower edge of the sample holder
is raised with respect to the microscope stage by about
$10^\circ$. The texture is tangential
counterclockwise.}\label{fig:tilt}
\end{figure}

To distinguish among these cases, the lower edge of the sample
holder is lifted producing a small tilt with respect to the
microscope stage. After the floating islands move to the edge of
the film due to gravity, their appearance changes to that shown in
Fig.~\ref{fig:tilt}.  The increased intensity on the right half of
the islands indicates that their c-director texture is tangential
counterclockwise, since in that case, the added tilt increases the
molecular tilt, relative to the propagation direction of the
light, and thus the apparent birefringence, on the right, and
decreases it on the left of the center.  All the +1 central defect
islands, as shown in Fig.~\ref{fig:bend}, have this ``pure bend''
texture throughout the island.

\begin{figure}
\centering
\includegraphics[width=2.5in]{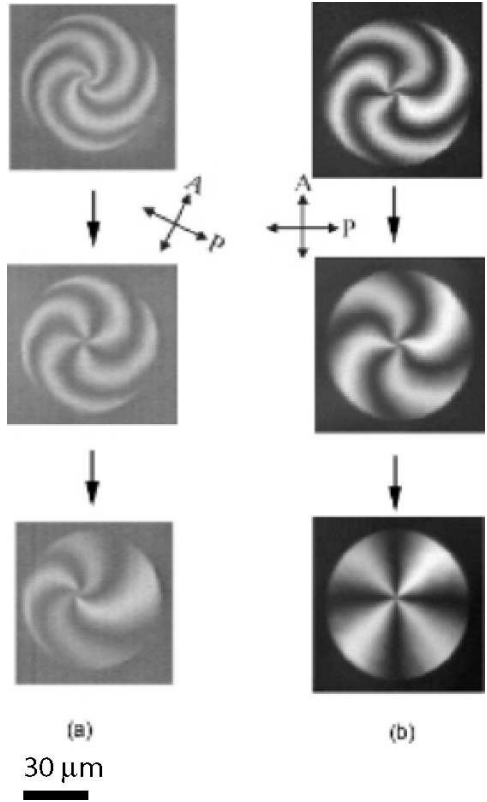}
\caption{The textures after blowing on islands with pure bend
texture. (a) A pure bend island switches to a 'simple spiral' with
radial c-director at the core. (b) A pure bend island exhibits a
transient (unstable) reversing spiral texture, when the core
boundary condition is unchanged.}\label{fig:simple}
\end{figure}

To explore the stability of pure bend islands, which are
apparently at equilibrium, we blew on the film with a small jet of
gas. The resulting swirling often induced two kinds of distinct
changes from a pure bend texture. In all cases the boundary
condition at the outer edge of the island remained tangential
counterclockwise.  However, the boundary condition at the smoke
particle at the core of the point defect would (a) sometimes
change from tangential to approximately radial, or (b) remain
tangential. Right after blowing, there was often a very tightly
wound spiral in the texture, which slowly relaxed to equilibrium,
as shown in Fig.~\ref{fig:simple}. Over the course of minutes, as
the c-director relaxes, in case (a) this texture equilibrates to a
final state with roughly $\pm\pi/2$ rotation from radial at the
core to tangential at the outer boundary (Bottom of
Fig.~\ref{fig:simple}(a)). We call this spiral texture a ``simple
spiral''. We have observed the simple spirals with radial inward
or outward boundary conditions at the core, and right or left
handed spirals. Neither the sense of the spiral nor the sign of
the radial boundary condition is meaningful, since observing the
sample from the other side would clearly reverse both these
characteristics. There is a size-dependence in this textural
transformation; pure bend islands of a small size are hardly ever
transformed into simple spirals. This indicates that there is only
a small interaction energy between the molecular orientation and
the smoke particle, since the pure bend texture is being
stabilized by the torque transmitted from the outer island
boundary to the defect core.

However the other spiral texture resulting from blowing, case (b),
which maintains its tangential boundary condition at the core,
which we call a 'reversing spiral', goes back to the pure bend
texture, as shown in Fig.~\ref{fig:simple}(b), when it started
from an apparently stable pure bend texture.

\begin{figure}
\centering
\includegraphics[width=2.5in]{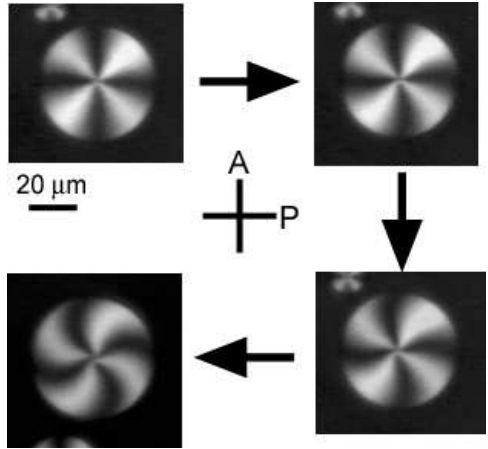}
\caption{Transformation of a pure bend island into a reversing
spiral island. It took around 30 min.}\label{fig:reversing}
\end{figure}

Remarkably, as shown in Fig.~\ref{fig:reversing}, as a stable pure
bend island grows, at some critical size the c-director texture
spontaneously starts to evolve into a reversing spiral.  This
transition is driven by a competition between the bend  and splay
elastic energies. If the splay elastic constant is less than the
bend constant, this enables the transformation from a pure bend
texture to a reversing spiral in a large enough island. The time
it takes to reach equilibrium depends on the thickness of the
island. The thinner the island, the less time. It took 30 minutes
for the pure bend island in Fig.~\ref{fig:reversing} to reach
equilibrium.

\section{Linear stability analysis}
To explore the stability of the texture in the islands, we employ
linear stability analysis. To find the equilibrium configuration
of $\hat{c}$ in the islands, we should minimize the following free
energy consisting of an integral of the elastic energy density
over the area of the island, and the edge energy at the sample
boundaries.
\begin{equation}\label{eq:free-energy}
F = \int[K_{s}(\nabla \cdot \hat{c})^{2}+ K_{b}(\nabla\times
\hat{c})^{2}]dA + \oint \sigma(\phi)dl.
\end{equation}
$K_s$ and $K_b$ are the two dimensional splay and bend curvature
elastic constants, respectively. $\phi$ is defined as the angle
between $\hat{c}$ and the outward radial vector $\hat{r}$.
$\sigma(\phi)$ is the anisotropic energy per unit length of edge.
Note that it is a periodic function,
$\sigma(\phi-2\pi)=\sigma(\phi)$. Generally it is written as
\begin{equation}
\sigma(\phi)=\sigma_{0} + \sum_{n=1}^{\infty} [a_{n}\cos(n\phi) +
c_{n}\sin(n\phi)],\label{eq:surface}
\end{equation}
in which $\sigma_{0}$ is an isotropic term
\cite{pettey,surface_energy}. In the areal free energy density, the
possible linear terms, $\nabla \cdot \hat{c}$ and $\nabla \times
\hat{c}$, are converted to line energies by the divergence theorem
and Stoke's theorem. Therefore the first order terms $a_1\cos{\phi}$
and $c_1\sin{\phi}$ in Eq.~(\ref{eq:surface}) arise from a
spontaneous splay($\nabla\cdot\hat{c}$) and a spontaneous
bend($\nabla\times\hat{c}$) respectively. However in a free-standing
film, which is physically the same when viewed from the top or
bottom, no spontaneous splay of $\hat{c}$ is possible; for films
floating on a liquid this term is allowed \cite{polar} by the polar
symmetry of the sample. So $a_{1}$ must vanish. $c_{1}$ arises
directly from the spontaneous bending (in our case, left turning)
produced by the molecular chirality in the SmC*, and it is directly
responsible for the minimum energy textures having a
counterclockwise tangential boundary condition at the outer edge of
all the islands we observed. We include in $\sigma(\phi)$ only the
lowest terms for the line energy,
\begin{equation}
\sigma(\phi)=\sigma_{0}+a_2\cos{2\phi}+c_{1}\sin{\phi}.
\end{equation}

For the islands we have created by a heterogeneous nucleation with
a strong anchoring tangential boundary condition at $r=R$ (outer
radius of island) and a weak anchoring boundary at $r=\epsilon$
(radius of the particle at the center), the line energy is
\begin{equation}
\oint\sigma(\phi)dl=2\pi R\sigma(\frac{\pi}{2})+2\pi
\epsilon\sigma(\phi(\epsilon)).\label{eq:sigma}
\end{equation}
The first term on the right side is the line energy for the outer
boundary and the second is for the inner boundary. The total free
energy for a island in Eq.~(\ref{eq:free-energy}) can be rewritten
as the sum of the areal elastic free energy and the line energy of
Eq.~(\ref{eq:sigma}) in terms of $x\equiv \ln(r/\epsilon)$.
\begin{eqnarray}
\nonumber F &=& \pi K \int_{0}^{x_{0}}\Big[(1 + \mu
\cos2\phi)\Big(\frac{d\phi}{dx}\Big)^{2}
+2\mu \sin2\phi \frac{d\phi}{dx} \\
&&+(1-\mu \cos2\phi)\Big]dx +
2\pi\epsilon\sigma(\phi(0)),\label{eq:energy}
\end{eqnarray}
where,
\begin{eqnarray}
\nonumber &&x_0 \equiv ln(R/\epsilon), \\
\nonumber &&K \equiv {\frac{K_{s}+K_{b}}{2}}, \\
\nonumber &&\mu \equiv {\frac{-K_{s}+K_{b}}{K_{s}+K_{b}}}.
\end{eqnarray}
The constant term $2\pi R\sigma(\pi/2)$ in the total free energy
is ignored, because it has no effect on $\phi(x)$. Here we assume
that the radial configuration of the $\hat{c}$ director, $\phi=0$
or $\phi=\pi$, in the areal free energy is a state for minimum
energy, based on our observation that $\hat{c}$ tends to rotate
toward a radial orientation in large enough islands. Hence
$K_{s}<K_{b}$, which means $\mu$ ranges from 0 to less than 1.

The equilibrium condition for $\phi(x)$ from the Euler-Lagrange
equation derived from Eq.~(\ref{eq:energy}) is
\begin{equation}
(1+\mu \cos2\phi){\frac{d^{2}\phi}{dx^{2}}}-\mu \sin2\phi
\Big({\frac{d\phi}{dx}}\Big)^{2}-\mu \sin2\phi =0.\label{eq:euler}
\end{equation}
The boundary condition at $x=0$ obtained from $(\frac{\partial
\eta(\phi)}{\partial \phi}-\frac{\partial f}{\partial
\phi'})\Big|_{x=0}=0$ is
\begin{equation}
K\Big[(1+\mu \cos2\phi)\Big({\frac{d\phi}{dx}}\Big)+\mu \sin2\phi
\Big]\Big|_{x=0}=\epsilon{\frac{d \sigma}{d
\phi}}\Big|_{x=0}\label{eq:torque},
\end{equation}
where $f$ and $\eta$ are the areal and edge free energy densities.

We found $\phi(x)$ numerically, using strong anchoring at the
outer boundary, $\phi(x_{0}) = \pi / 2$, as required by our
observation. However, depending on the choice of  parameters,
there might be one or more numerical solutions. Linear stability
analysis is used to determine if these solutions define local
minima for the free energy. For the analysis, we need to
manipulate the total energy in Eq.~(\ref{eq:energy}). Defining
\begin{eqnarray}
&&A(\phi)\equiv 1+\mu \cos{2\phi},\\
&&B(\phi)\equiv 2\mu \sin{2\phi},\\
&&C(\phi)\equiv 1-\mu \cos{2\phi},
\end{eqnarray}
Eq.~(\ref{eq:energy}) can be rewritten as
\begin{eqnarray}\label{F}
F & = & \pi K
\int_{0}^{x_{0}}dx\Big[A(\phi)\Big(\frac{d\phi}{dx}\Big)^{2}+
B(\phi)\frac{d\phi}{dx}+C(\phi)\Big] \nonumber \\
&& + \; 2\pi\epsilon\sigma(\phi(0)).
\end{eqnarray}
Let's start with the second term in the integral.
\begin{eqnarray}
\nonumber \pi K\int^{x_0}_{0}B(\phi)\frac{d\phi}{dx}dx
&=&\pi K\int^{\phi{(x_0)}}_{\phi(0)}B(\phi)d\phi\\
\nonumber &=&\pi K\mu(\cos2\phi(0)-\cos2\phi(x_0))\\
&=& \pi K\mu(\cos2\phi(0)+1).
\end{eqnarray}
We rewrite Eq.~(\ref{F}).
\begin{eqnarray}
F & = & \pi
K\int^{x_0}_{0}\Big[A(\phi)\Big(\frac{d\phi}{dx}\Big)^2+C(\phi)\Big]dx
\nonumber \\
&& + \; \frac{1}{2}\pi KS(\phi(0)),
\end{eqnarray}
where $\frac{1}{2}\pi KS(\phi(0))\equiv 2\pi\epsilon
\sigma(\phi(0))+\pi K \mu \cos(2\phi(0))+\pi K \mu$. We next
incorporate the line energy into the integral and symmetrize the
problem around $x=0$, obtaining
\begin{equation}
F=\frac{\pi
K}{2}\int^{x_0}_{-x_0}\Big[A(\phi)\Big(\frac{d\phi}{dx}\Big)^2
+C(\phi)+\delta(x)S(\phi(x))\Big]dx,\label{eq:deltaenergy}
\end{equation}
in which $\delta(x)$ is a Dirac delta function.

Now we expand the above free energy to second order in a small
fluctuation $\varphi=\phi-\phi_{0}$, in which $\phi_{0}$ is a
solution that makes the total free energy $F$ an extremum. The
boundary conditions are $\varphi(-x_0)=\varphi(x_0)=0$ due to the
strong anchoring. The expanded free energy $\delta F$ is
\begin{eqnarray}
\nonumber \delta F&=&\frac{\pi K}{2}
\int^{x_0}_{-x_0}\bigg\{\Big[A(\phi_0)+A'(\phi_0)\varphi+
\frac{1}{2}A''(\phi_0)\varphi^2\Big] \nonumber \\
 &&\times \Big[\Big(\frac{d\phi_0}{dx}\Big)^2
+ \; 2\frac{d\phi_0}{dx}\frac{d\varphi}{dx}
+ \; \Big(\frac{d\varphi}{dx}\Big)^2\Big]\nonumber \\
&&+
\;\Big[C(\phi_0)+C'(\phi_0)\varphi
+\frac{1}{2}C''(\phi_0)\varphi^2\Big]\\
&&+\delta(x)\Big(S(\phi_0)+S'(\phi_0)\varphi
+\frac{1}{2}S''(\phi_0)\varphi^2\Big)\bigg\}dx,\nonumber
\end{eqnarray}
where the notation $'$ and $''$ denote first and second
derivatives with respect to $\phi$.

Since we are expanding around an extremum solution, the first
functional derivative $\delta F_1=0$. As a result, the lowest term
is the second order derivative of the free energy,
\begin{eqnarray}\label{Ftwo}
\nonumber \delta F_2&=&\frac{\pi K}{2}
\int^{x_0}_{-x_0}\Big[\frac{1}{2}A''(\phi_0)
\Big(\frac{d\phi_0}{dx}\Big)^2\varphi^2
+A(\phi_0)\Big(\frac{d\varphi}{dx}\Big)^2\\
&&+2A'(\phi_0)\frac{d\phi_0}{dx}\varphi
\frac{d\varphi}{dx}+\frac{1}{2}C''(\phi_0)\varphi^2 \nonumber
\\
&&+\delta (x)\frac{1}{2}S''(\phi_0)\varphi^2\Big]dx.
\end{eqnarray}
We next convert the second derivative terms to a symmetric form
by integration by parts.
\begin{eqnarray}\label{A}
\nonumber A(\phi_0)\Big(\frac{d\varphi}{dx}\Big)^2 & \rightarrow &
-A'(\phi_0)\frac{d\phi_0}{dx}\frac{d\varphi}{dx}\varphi
-A(\phi_0)\frac{d^2\varphi}{dx^2}\varphi,\\
2A'(\phi_0)\frac{d\phi_0}{dx}\varphi \frac{d\varphi}{dx} &
\rightarrow & -A''(\phi_0)\Big(\frac{d\phi_0}{dx}\Big)^2 \varphi^2
\nonumber \\ && -  A'(\phi_0)\frac{d^2\phi_0}{dx^2}\varphi^2.
\end{eqnarray}
After substituting Eq.~(\ref{A}) into Eq.~(\ref{Ftwo}), we obtain
\begin{eqnarray} \nonumber \delta F_2 &=& \frac{\pi K}{2}
\; \int_{-x_{0}}^{x_{0}}\varphi
\Big[-\frac{d}{dx}\Big(A(\phi_{0})\frac{d}{dx}\Big) \nonumber \\
&& - \frac{1}{2}A^{''}(\phi_{0})\Big(\frac{d\phi_{0}}{dx}\Big)^{2}
-A^{'}(\phi_{0})\frac{d^{2}\phi_{0}}{dx^{2}}\nonumber
\\
&& + \; \frac{1}{2}\; C^{''}(\phi_{0}) +\frac{1}{2} \;
\delta(x)S^{''}(\phi_{0}(0))\Big]\varphi dx.
\end{eqnarray}
$\delta F_2$ has the form of
\begin{equation}
\delta F_2=\frac{\pi K}{2} \int^{x_0}_{-x_0}[\varphi
\textit{L}\varphi]dx,
\end{equation}
where $\textit{L}$ is the operator
$-\frac{d}{dx}(A(\phi_0)\frac{d}{dx})+V(x)$ and $V(x)$ is
completely determined by the solution $\phi_0(x)$. This
symmetrized second derivative term in $\textit{L}$ is now
Hermitian or self adjoint, as is $V(x)$. The operator $\textit{L}$
has an infinite set of eigenfunctions which form a complete
orthogonal set of functions for constructing all suitable
functions satisfying the boundary conditions on the
integral.\cite{mandf} The eigenfunction with the lowest eigenvalue
does not cross zero between the $\pm x_0$ end points, while
sequentially, each higher eigenvalue corresponds to an
eigenfunction with one more zero crossing. Let's expand a test
function $\varphi$ in the eigenfunctions of $\textit{L}$,
$\varphi=\sum{a_i\varphi_i}$, in which $\varphi_i$ is normalized
so that
\begin{equation}
\int^{x_0}_{-x_0}\varphi_i^2 dx=1,
\end{equation}
and each eigenfunction $\varphi_i$ has an eigenvalue $C_i$. Then
\begin{equation}
\delta F_2=\frac{\pi K}{2}\int^{x_0}_{-x_0}\varphi \textit{L}
\varphi dx=\frac{\pi K}{2}\sum{a_i^2C_i}
\end{equation}
If the lowest eigenvalue $C_1>0$, $\delta F_2$ is positive, and
$\phi_0$ minimizes the free energy with respect to small
variations. If $C_1<0$, then $\varphi_1$ clearly lowers the free
energy and $\phi_0$ is not a minimizing function. To test a
solution $\phi_0$ we find the eigenfunction of the operator
$\textit{L}$, which does not cross zero between the endpoints, and
determine its eigenvalue. A negative eigenvalue indicates
instability.

\section{Results}
We have described the procedures to obtain the equilibrium textures
and to examine their stability against infinitesimal fluctuations to
understand what we observe, which is that there can be more than one
stable or metastable solution,  for the texture of an island. To
find minima of the free energy, we numerically solve the
Euler-Lagrange equation Eq.~(\ref{eq:euler}) with the boundary
conditions,  Eq.~(\ref{eq:torque}) at the core and strong anchoring
$\phi=\pi/2$ at the outer boundary. We adopt the shooting method, in
which starting from the outer boundary with $\phi=\pi/2$, we vary
the slope of $\phi$, and integrate to the inner boundary. The
discrepancy from the desired inner boundary condition is used to
adjust the initial slope, repeating until the solution is obtained.

\begin{figure}
\centering
\includegraphics[width=3.5in]{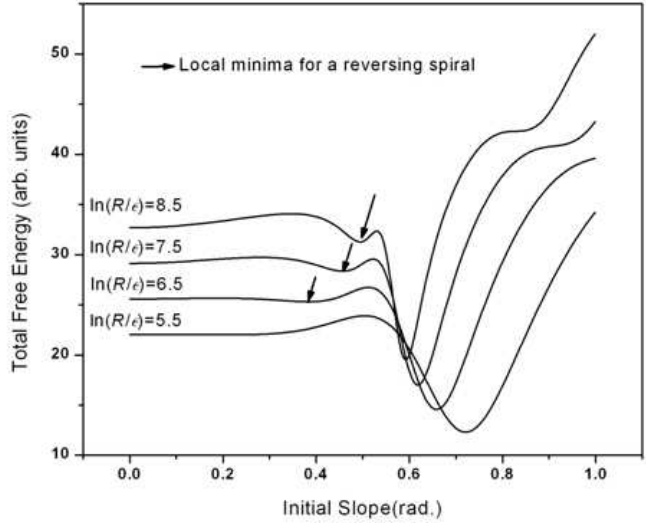}
\caption[Total free energy vs. initial slope]{ A plot showing the
minima of the free energy of an island as a function of the
initial slope of $\phi$ at the outer boundary for the following
parameters, $K=1$, $\mu=0.13$, $\epsilon c_1=1.0$, and $\epsilon
a_2=0.6$. At initial slope 0, the free energy is for a pure bend
island.}\label{fig:shooting}
\end{figure}

In Fig.~\ref{fig:shooting}, we show the free energy of functions
$\phi$ as the slope at the outer boundary is varied.  The local
minima represent solutions of the Euler-Lagrange equation that
satisfy the boundary condition at the inner boundary.  For the four
island sizes shown, compare the free energy of the pure bend island,
with initial slope zero, to that of the other local minima.  The
global minimum at large initial slope represents a simple spiral,
while the local minimum that appears for island size just above
$\ln{(R/\epsilon)}=5.5$ is a reversing spiral. Actually many local
minima can exist at larger initial slope, but they have much higher
free energy than the free energy of the pure bend island.  The
linear stability analysis of the pure bend texture shows that once
the reversing spiral solution appears, the pure bend texture is
unstable.

\begin{figure}
\centering
\includegraphics[width=2.5in]{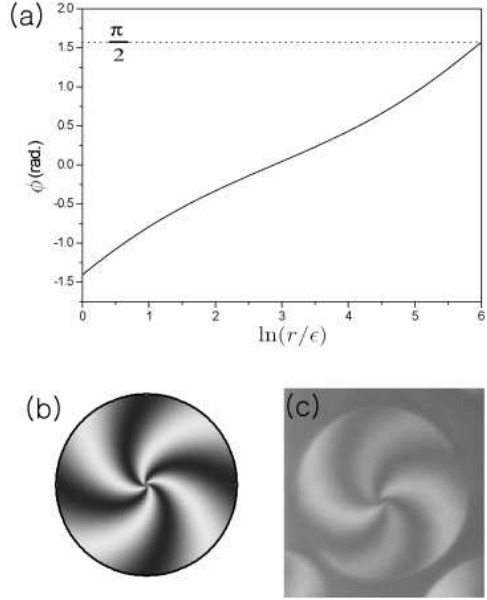}
\caption[Stable solution for a simple spiral.]{Stable solution for
a simple spiral. $K$=1, $\mu$=0.13, $\epsilon c_1$=1.0, and
$\epsilon a_2$=0.6 at $\ln(R/\epsilon)$=6.0. (b) Texture of the
solution as seen between crossed polarizers. (c) A simple spiral
we observed, after blowing on the
sample.}\label{fig:simple_fitting}
\end{figure}

An example of a stable simple spiral is illustrated in
Fig.~\ref{fig:simple_fitting}. In
Fig.~\ref{fig:simple_fitting}(a), $\phi$ varies monotonically from
the core to the outer boundary. This is an illustration of the
lowest energy configuration for $\mu$=0.13, $K$=1, $\epsilon
c_{1}$=1.0, and $\epsilon a_{2}$=0.6 at $\ln(R/\epsilon)$=6.0.
Fig.~\ref{fig:simple_fitting}(b) represents the appearance of the
island, as seen between crossed polarizers, for which $\phi(x)$ is
shown in Fig.~\ref{fig:simple_fitting}(a). The island in
Fig.~\ref{fig:simple_fitting}(c) is a typical simple spiral we
created by blowing.  Note that in the case calculated here, the
stable boundary condition at the core is not radial but almost
reversed tangential, but the part of the solution with values of
$\phi <0$ is so close to the core that it is essentially invisible
at optical resolution, since the radius of the island is only on
the order of 20 micrometers. The important qualitative observation
is that $\phi$ increases monotonically from the core. In fact we
do not know the precise core size, just that it is sub-visible.

\begin{figure}
\centering
\includegraphics[width=2.5in]{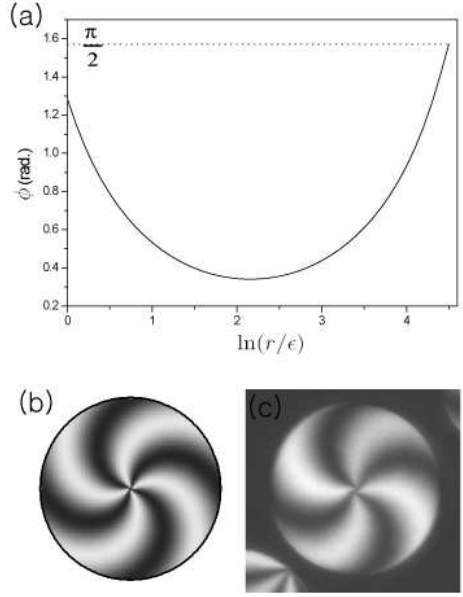}
\caption[Stable solution for a reversing spiral.]{Stable solution
for a reversing spiral. $K$=1, $\mu$=0.6, $\epsilon c_1$=1.0, and
$\epsilon a_2$=0.5 at $\ln(R/\epsilon)$=4.5. (b) The texture seen
between crossed polarizers. (c) A typical reversing spiral island.
}\label{fig:reversing_fitting}
\end{figure}

In the reversing spiral seen in Fig.~\ref{fig:reversing_fitting},
$\phi$ rotates in one direction from the inner boundary until
reaching a minimum angle, and then rotates in the other direction
to match the outer boundary condition. This configuration does not
represent an absolute minimum of the free energy, but is locally
stable for $\mu$=0.6, $K$=1, $\ln(R/\epsilon)$=4.5, $\epsilon
c_{1}$=1.0, and $\epsilon a_{2}$=0.5.
Fig.~\ref{fig:reversing_fitting}(b) shows the appearance of the
island as seen between crossed polarizers, for the solution
$\phi(x)$ shown in Fig.~\ref{fig:reversing_fitting}(a).
Fig.~\ref{fig:reversing_fitting}(c) is the texture of an observed
reversing spiral.  Again, the detailed texture near the core is
unresolved, but qualitatively it is clear that the spiral reverses
curvature at very small radius.

\begin{figure}
\centering
\includegraphics[width=3.5in]{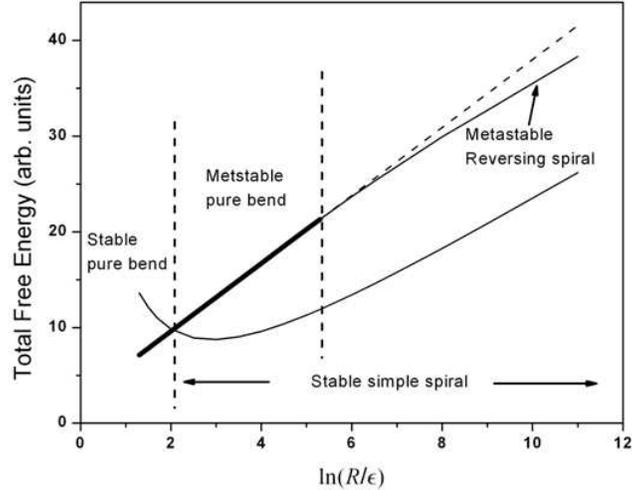}
\caption[Phase diagram for an island.]{State diagram for an
island. $K$=1, $\mu$=0.13, $\epsilon c_1$=1.0, and $\epsilon
a_2$=0.6.}\label{fig:phasediagram}
\end{figure}

Fig.~\ref{fig:phasediagram} is a state diagram for an island, as a
function of its radius, obtained by numerical calculation, where
the simulation parameters are $K$=1, $\mu$=0.13, $\epsilon
c_1$=1.0, and $\epsilon a_2$=0.6. At small radius, the energy for
a simple spiral is much higher than the energy of a pure bend
texture. This explains why we can't observe the simple spirals in
islands of small size. Above a certain size, the pure bend island
is only metastable, but there is a sizable energy barrier for
transition to the simple spiral.  However, above a second critical
size, the pure bend island becomes unstable relative to a
reversing spiral with a lower energy, without any energy barrier.
The transition size is sensitive to $K_{s}$ and $K_{b}$. If
$K_{s}$ and $K_{b}$ are the same, the size of the transition
diverges. The bigger $\mu$, the smaller the size for this second
order transition.

For island radius larger than the radius for the crossing of the
free energies of the pure bend texture and the simple spiral,
around $\ln (R/\epsilon)=2$ in this case, the pure bend texture
can be transformed into a simple spiral as a first order
transition. We think blowing on the sample makes this possible in
our experiments, because the highly curled transient spiral state
it produces provides the torque needed to change the inner
boundary condition.  The spiral then relaxes into the stable
simple spiral state.

\section{Discussion}

We have observed the textures of circular islands of a
ferroelectric smectic C* liquid crystal, especially studying first
and second order transformations of textures as a function of
island size.  First our basic observations confirm the fact that
the chirality of the smectic C* phase induces a preferred bend
curvature in the plane of the smectic layers, leading in the case
of the materials we studies, to a general left turning bend
texture as their ground state.  Second, using an elastic model in
which the difference of elastic constants, $K_b-K_s$, provides a
driving force, and the anchoring of the c-director at the core of
the central disclination is weak, we construct a plausible state
diagram that agrees with our observations.  The circular geometry
of the island, combined with the strong tangential anchoring of
the c-director at its outer boundary, demand a curved texture,
which makes the difference between the splay and bend curvature
elastic constants crucial. The continuous transition from the pure
bend island to the reversing spiral is completely analogous
mathematically to the behavior of a plane parallel sheet of
nematic with tangential boundary conditions, in a perpendicular
magnetic field. As the sheet grows thicker, at a critical
thickness, a continuous Frederiks transition occurs, in which the
director rotates toward the external field direction. Here, that
effective field direction is radial, converting bend into splay,
and lowering the energy.

In studying different materials, using the textural transformation
from pure bend to reversing spiral as a diagnostic, we saw a
strong correlation between the spontaneous electrical polarization
of the ferroelectric phase and the apparent difference between the
bend and splay elastic constants; high polarization materials
matched our model of a material with a high bend elastic constant.
This result correlates well with the fact that bend curvature of
the c-director produces divergence of the spontaneous
polarization, resulting in space charges which interact to
increase the free energy of the bent state.  In low frequency, or
quasi-static, conditions, the resulting space charge is partially
screened by free charges in the material, reducing the otherwise
long range electrostatic interactions to an effective local
contribution to the free energy, which can be expressed as an
increase in the effective bend elastic constant.\cite{okano} The
fact that our state diagram and the textures we calculate from the
elastic model agree well with our observations leads us to
hypothesize that we have successfully approximated the
electrostatic interactions and energy by using a large bend
elastic constant $K_b$ to account for the behavior of high
polarization materials. The high value of $K_b$ pushes the
critical island size for the appearance of the reversing spiral to
very small values, as one would expect qualitatively, and as we do
observe for high polarization materials. To explore this
hypothesis further, we made independent studies of the elastic
constants and the interactions of bend distortions with free
charges in free standing SmC* films, comparing high frequency and
quasi-static behavior.  To accomplish this, we performed light
scattering experiments on c-director fluctuation dynamics, with
electric field quenching of fluctuations, with analysis including
both bend-induced polarization charge and conduction charge
screening. They will be reported elsewhere.


\begin{thebibliography}{99}
\bibitem{equilibriumsize} R.~B.Meyer, D. Konovalov, I. Kraus and J.-B. Lee,
Mol. Cryst. Liq. Cryst. {\bf 364} 123-131 (2001).
\bibitem{polar} I. Kraus and R.~B. Meyer, Phys. Rev. Lett.
$\textbf{82}$, 3815 (1999).
\bibitem{muzny} C.~D. Muzny and N.~A. Clark, Phys. Rev. Lett.
$\textbf{68}$, 804 (1992).
\bibitem{maclennan} J.~E. Maclennan, U. Sohling, N.~A. Clark, and
M. Seul, Phys. Rev. E $\textbf{49}$, 3207 (1994).
\bibitem{pettey} D. Pettey and T.~C. Lubensky, Phys. Rev. E
$\textbf{59}$, 1834 (1999).
\bibitem{rivier} S. Riviere and J. Meunier, Phys. Rev. Lett.
$\textbf{74}$, 2495 (1995).
\bibitem{fang} J. Fang, E. Teer, C.~M. Knobler, K.-K. Loh, and J.
Rudnick, Phys. Rev. E $\textbf{56}$, 1859 (1997).
\bibitem{sethna} S.~A. Langer and J.~P. Sethna, Phys. Rev. A
\textbf{34}, 5035 (1986).
\bibitem{loh} K.-K. Loh, I. Kraus, and R.~
B. Meyer, Phys. Rev. E \textbf{62}, 5115 (2000).
\bibitem{jb} J.-B. Lee, R. B. Meyer, and R. A. Pelcovits, to be
published. J.-B. Lee, Ph.D. thesis, Brandeis University, (2004).
\bibitem{surface_energy} J. Rudnick and R. Bruinsma, Phys. Rev.
Lett. \textbf{74}, 2491 (1995).
\bibitem{mandf} For a clear discussion of Sturm Liouville theory
see Morse and Feshbach, {\it Methods of Theoretical Physics},
McGraw-Hill, 1953.
\bibitem{okano} K. Okano, Jpn. J. Appl. Phys. $\textbf{25}$, L846
(1986).
\end{thebibliography}
\end{document}